\documentclass[useAMS,usenatbib]{mn2e}
\usepackage{amsmath,amstext,amsgen,amsbsy,amsopn,amsfonts,theorem}

\usepackage[pdftex]{graphicx}
\usepackage{xspace}
\usepackage{amsmath}
\usepackage{hyperref}
\usepackage{framed}
\usepackage{txfonts}
\usepackage{epstopdf}
\usepackage{gensymb}

\usepackage{subcaption}

\usepackage[normalem]{ulem}

\makeatletter
\def\mr@ignsp#1 {\ifx\:#1\@empty\else #1\expandafter\mr@ignsp\fi}%
\newcommand{\multiref}[1]{\begingroup
\xdef\mr@no@sparg{\expandafter\mr@ignsp#1 \: }%
\def\mr@comma{}%
\@for\mr@refs:=\mr@no@sparg\do{\mr@comma\def\mr@comma{,}\ref{\mr@refs}}%
\endgroup}
\makeatother

\newcommand{\hypref}[2]{\ifx\href\asklfhas #2\else\href{#1}{#2}\fi}
\newcommand{\Secref}[1]{Section~\multiref{#1}}

\newcommand{\Appref}[1]{Appendix~\multiref{#1}}

\newcommand{\Figref}[1]{Figure~\multiref{#1}}
\newcommand{\figref}[1]{Fig.~\multiref{#1}}
\renewcommand{\eqref}[1]{(\multiref{#1})}

\title[Hypervelocity stars in Gaia]{A hypervelocity star with a Magellanic origin}

\author[D. Erkal et al.]
  { Denis Erkal$^{1}$\thanks{d.erkal@surrey.ac.uk}, Douglas Boubert$^2$, Alessia Gualandris$^1$, N. Wyn Evans$^2$, Fabio Antonini$^1$ \\
  $^1$Department of Physics, University of Surrey, Guildford GU2 7XH, UK \\
  $^2$Institute of Astronomy, University of Cambridge, Madingley Road, CB3 0HA, Cambridge, UK
   }

\begin{document}

\label{firstpage}

\maketitle

\begin{abstract}
Using proper motion measurements from \textit{Gaia} DR2, we probe the origin of 26 previously known hypervelocity stars (HVSs) around the Milky Way. We find that a significant fraction of these stars have a high probability of originating close to the Milky Way centre, but there is one obvious outlier. HVS3 is highly likely to be coming almost from the centre of the Large Magellanic Cloud (LMC). During its closest approach, $21.1^{+6.1}_{-4.6}$ Myr ago, it had a relative velocity of $870^{+69}_{-66}$ kms$^{-1}$ with respect to the LMC. This large kick velocity is only consistent with the Hills mechanism, requiring a massive black hole at the centre of the LMC. This provides strong direct evidence that the LMC itself harbours a massive black hole of at least $4\times 10^3 -10^4 M_\odot$. 
\end{abstract}


\section{Introduction}

A hypervelocity star is one whose velocity is sufficiently fast that it will escape the gravitational pull of the Milky Way. The existence of this class of star was theorised by \citet{hills_hyper-velocity_1988}. If a binary star is disrupted by the massive black hole Sgr A*, then one of the stars could be ejected from the Galactic centre at more than $1000\;\mathrm{km}\;\mathrm{s}^{-1}$. The first hypervelocity star (termed `HVS1') was discovered by \citet{brown_discovery_2005} and interpreted as a vindication of the prediction. HVS1 is a $3\;\mathrm{M}_{\odot}$ B star at a distance of $107\pm15\;\mathrm{kpc}$ from the Sun and its extreme radial velocity $831.1\pm5.7\; \mathrm{km}\;\mathrm{s}^{-1}$ proves that it is unbound from the Galaxy \citep{brown_mmt_2014}. The second hypervelocity star (HVS2) was discovered by \citet{hirsch_us_2005} and is rather different: the sdO star US708 lies at only $8.5\pm1.0\;\mathrm{kpc}$ and has a heliocentric velocity of $917\pm7\;\mathrm{km}\;\mathrm{s}^{-1}$ \citep{geier_fastest_2015}. The third hypervelocity star (HE 0437-5439 or HVS3) was discovered by \citet{edelmann_he_2005} and was similar to HVS1 rather than HVS2, being an $8\mathrm{M}_{\odot}$ B star at a distance of $61\;\mathrm{kpc}$. Brown and collaborators subsequently conducted a targeted search for hypervelocity stars and identified a further twenty B hypervelocity stars in the halo of the Milky Way which were similar to HVS1 \citep[HVS4 through HVS24 excluding HVS11][]{brown_hypervelocity_2006,brown_hypervelocity_2007,brown_mmt_2014}. Further B type hypervelocity star discoveries include HD 271791 \citep{heber_b-type_2008}, HIP 60350 \citep{irrgang_nature_2010} and LAMOST-HVS1 through LAMOST-HVS3 \citep{zheng_first_2014,huang_discovery_2017}.

There have been numerous claimed late-type hypervelocity stars within a few kiloparsecs of the Sun \citep[see][and references therein]{brown_hypervelocity_2015}. However, these candidates are much less certain than the early-type B hypervelocity stars in the halo. B stars live short lives and thus their extreme radial velocities are corroborated by their need to travel that rapidly in order to survive the trip from regions of active star formation to their location in the halo. Late-type hypervelocity stars are often classed as HVSs based on large proper motions and there are several cases where their proper motions have proven erroneous. These candidates have recently been re-assessed by \citet{Bo18}  in the light of \textit{Gaia} DR2 and few have survived the purging.
For this reason, we focus on the origins of the confirmed B-type hypervelocity stars here.

\citet{brown_hypervelocity_2015} argues that an origin through the \citet{hills_hyper-velocity_1988} mechanism operating in the Galactic centre remains the only plausible origin of the hypervelocity stars if they originate in the Milky Way. However for HVS3, this assumption has been questioned before. \citet{edelmann_he_2005} noted that HVS3 is only $16.3\;\deg$ from the centre of the Large Magellanic Cloud (LMC) and that (neglecting the potential of the LMC and without measured proper motions) the star was consistent with being ejected from the LMC's centre $35\;\mathrm{Myr}$ ago at $600\;\mathrm{km}\;\mathrm{s}^{-1}$. An origin in the LMC solves the problem that HVS3 would not live long enough to survive the journey from the Galactic centre to its current location\citep{edelmann_he_2005}. \citet{gualandris_hvs3} found that the ejection of this star from the LMC at such a rapid velocity would require the \citet{hills_hyper-velocity_1988} mechanism to be operating, and that the required black hole would need to be at least $10^3\;\mathrm{M}_{\odot}$. \citet{brown_galactic_2010} used proper motions measured with the Hubble Space Telescope (HST) to argue that the star likely originated in the Milky Way. However, with additional HST data, \citet{brown_proper_2015} found that HVS3 could have come from either the Milky Way or the LMC. \citet{perets_dynamical_2009} suggests that the discrepancy between the flight time from the Galactic Centre and the stellar lifetime may be resolved by assuming that the star was ejected as a binary with hypervelocity and later merged with its companion due to internal processes. However, the ejection of hypervelocity binaries following the disruption of a stellar triple is extremely unlikely \citep{fragione_gual_2018}.

An origin in the LMC has recently been proposed for many of the other B hypervelocity stars. \citet{boubert_dipole_2016} showed if there is a massive black hole in the LMC then the Hills mechanism would cause stars to be ejected from the LMC, and that those stars would match the kinematic pattern of the observed hypervelocity stars. \citet{boubert_hvs_runaway} investigated the ejection of stars from the LMC by the supernova of their companion and showed that this more standard scenario could also explain the hypervelocity stars observed in the halo.

The European Space Agency's Gaia space telescope was launched in 2013 and on the 25$^{\mathrm{th}}$ April 2018 delivered its second date release \citep[Gaia DR2,][]{2016A&A...595A...1G,2018arXiv180409365G}  containing astrometry and photometry for 1,692,919,135 sources, based on the first 22 months of operation. This catalogue has parallaxes and proper motions for 1,331,909,727 sources, including for a majority of the hypervelocity stars considered in this work.

It is in this context that we investigate whether the B hypervelocity stars are more consistent with a Milky Way or LMC origin, using a set of proper motions synthesised from the literature (most notably, \citealp{brown_proper_2015} obtained HVS proper motions for HVS1-13) and from the newly released \textit{Gaia} DR2 (see Tab. \ref{tab:hvs}). In Section \ref{sec:methods}, we describe how we follow back in time the trajectories of the hypervelocity stars accounting for the Galactic and LMC potentials and the measurement errors. Section~\ref{sec:hvs3} discusses HVS3 in some detail, building the case that it has been ejected by a super-massive black hole at the centre of the Large Magellanic Cloud. In Section \ref{sec:testing}, we validate our method using mock data from the \citet{boubert_dipole_2016} simulations of stars ejected from the LMC's centre.

\begin{figure*}
\centering
\includegraphics[width=0.45\textwidth]{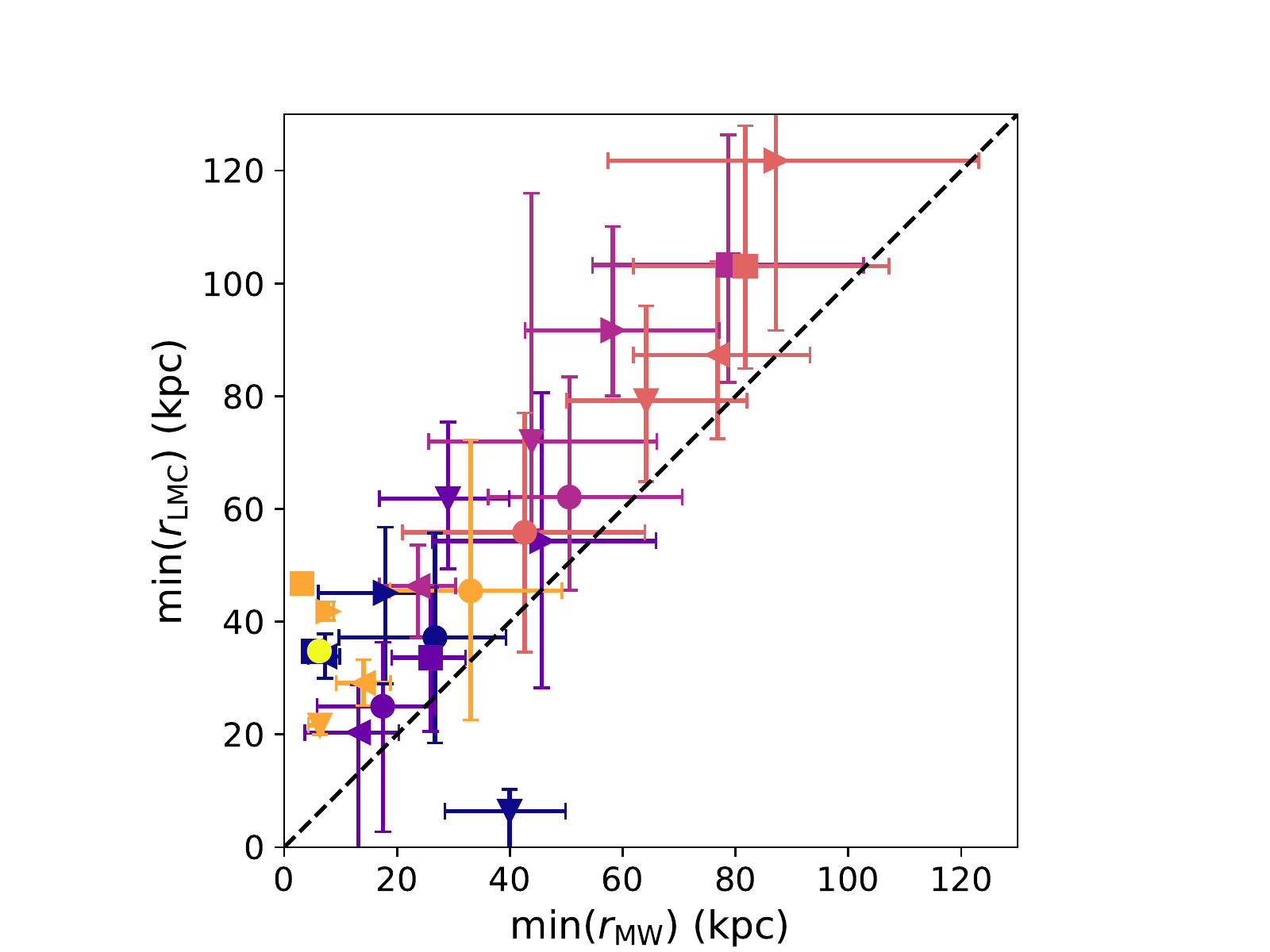}
\includegraphics[width=0.45\textwidth]{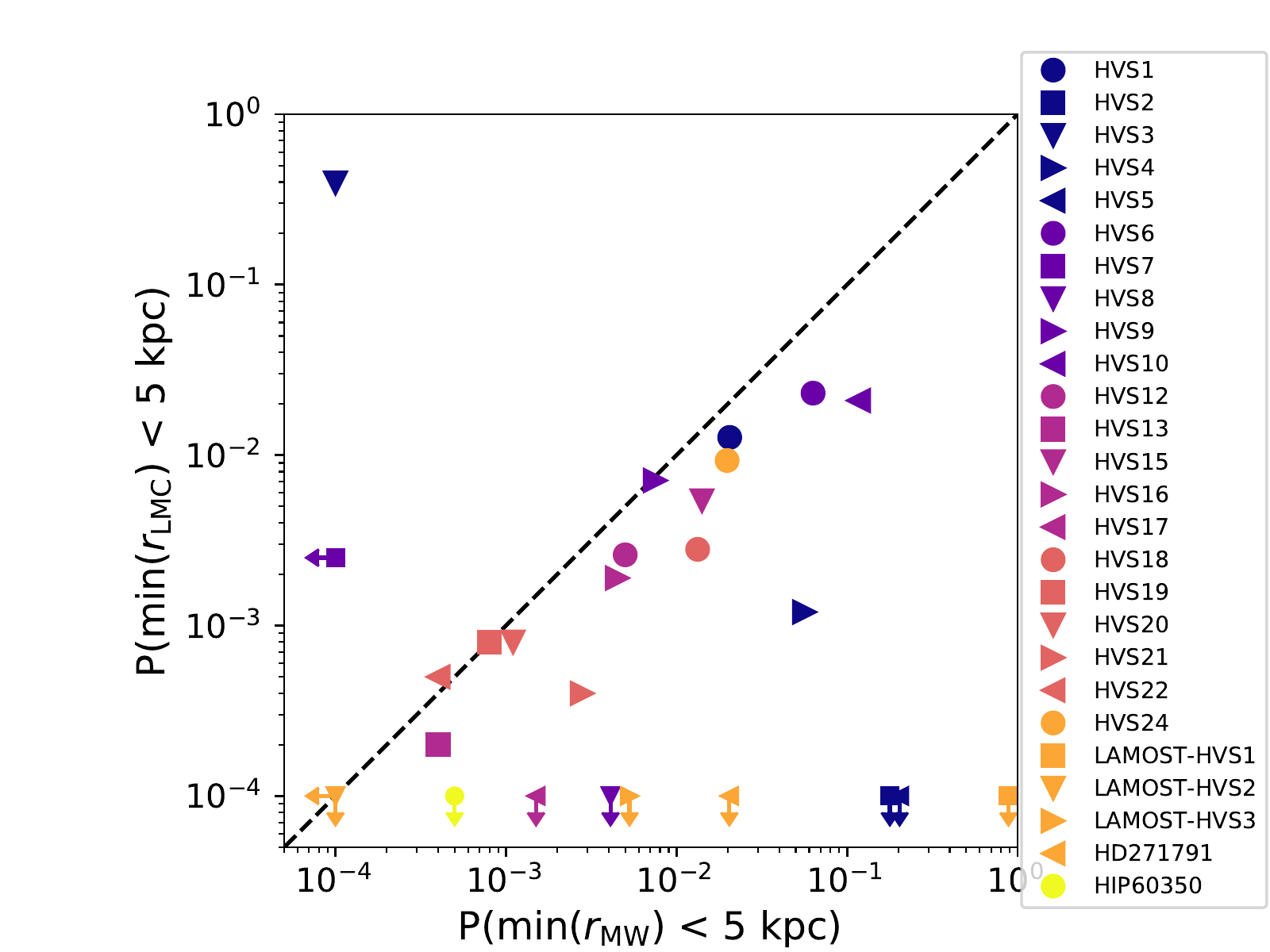}
\caption{\textbf{Left:} Minimum approach distance to the Milky Way and the LMC for the 26 HVSs in our sample. HVS3 stands out as being far more likely to have originated near the LMC than near the Milky Way. \textbf{Right:} Probability of passing within 5 kpc of the Milky Way and the LMC over the past 500 Myr. Again, HVS3 clearly stands out as being significantly more likely to originate from near the centre of the LMC ($p =  0.40$) than the MW ($p=10^{-4}$). } 
\label{fig:HVS_MW}
\end{figure*}

\section{Population of hypervelocity stars} \label{sec:population}

\subsection{Method}
\label{sec:methods}

In order to determine the origin of the HVSs, we rewind them in the combined presence of the gravitational field of the Milky Way and the LMC. For the Milky Way, we consider two potentials. First, we use \texttt{MWPotential2014} from \cite{galpy} which satisfies a number of observed constraints. This potential consists of an NFW dark matter halo \citep{nfw_1997},a Miyamoto-Nagai disk \citep{mn_disk}, and a power-law bulge with an exponential cutoff. When using this potential, we use an offset of $(U_\odot, V_\odot, W_\odot) = (-11.1, 24, 7.3)$ km s$^{-1}$ from \cite{schoenrich_etal_2010} and \cite{bovy_etal_vcirc} for the Sun's motion relative to the local circular velocity. We assume that the Sun is located at a distance of $8.3\pm0.1$ kpc from the Galactic centre. The second potential that we consider is the recent Milky Way model in \cite{pjm17}.

For the LMC, we choose a mass of $1.5\times 10^{11}M_\odot$. This is based on the several lines of reasoning which have suggested that the LMC has a mass on the order of $10^{11}M_\odot$ based on abundance matching \citep{behroozi_etal_2013,moster_etal_2013}, requiring that the SMC is bound to the LMC \citep{kallivayalil_etal_2013}, the plethora of dwarfs around the LMC \citep{jethwa_lmc_sats}, as well as accounting for the effect of the LMC on the timing argument between MW and M31 \citep{penarrubia_lmc}. The LMC is modelled as a Hernquist profile with a scale radius of 17.14 kpc. This profile satisfies the rotation curve measurements by \cite{vandermarel_lmc}. We note that we have performed the analysis with an LMC mass of $2\times 10^{10} M_\odot$ and we find similar results. We use an LMC distance of $49.97\pm1.126$ kpc \citep{pietrzynski_lmc_dist}, an LMC radial velocity of $262.2\pm0.3$ km/s \citep{vandermarel_lmc_rv}, and proper motions of $(\mu_\alpha \cos \delta ,\mu_\delta= (1.91\pm0.02,0.229\pm0.047)$ mas/yr \citep{kallivayalil_etal_2013}. We include dynamical friction from the Milky Way on the LMC using the prescription of \cite{hashimoto_etal_2003}.

\subsection{General Properties of the Population}

The properties of the HVSs are given in \Appref{sec:hvs_props} and come from a variety of sources. \textit{Gaia} DR2 has significantly improved the proper motions of many of the HVSs. With the updated proper motions in hand, we now integrate their orbits backwards in the Milky Way potential to determine the point of closest approach of the HVSs and the Milky Way. We choose to rewind the HVSs for 500 Myr which is significantly longer than their main sequence lifetimes; most of the stars are late B or early A type with the sole exception of HVS2 (also known as US708) which is a nearby sdO star with an evolutionary lifetime of $100\;\mathrm{Myr}$ \citep{hirsch_us_2005}. The rewinding is done by performing a Monte Carlo sampling of the 6d phase space coordinates of the HVS based on the observational uncertainties in distance, proper motion (accounting for the covariance in proper motions from \textit{Gaia} DR2), and radial velocity. We also sample the distance of the Sun from the Galactic centre as well as the 6d phase space coordinates of the LMC. For each realization of these quantities, we rewind each HVS for 500 Myr and determine its point of closest approach with the Milky Way and the LMC. We then record the distance, relative velocity, and lookback time to this close passage. In addition, we determine the location where each HVS passes through the plane of the Milky Way disk and the plane of the LMC disk \citep[as measured in][]{vandermarel_lmc}.

In the left panel of Figure~\ref{fig:HVS_MW}, we show the minimum approach distance to the Milky Way and LMC for the HVSs in the sample. Points on the top left of this figure have a closer approach to the Milky Way than the LMC and vice versa for the points on the bottom right. The right panel shows the probability of passing within 5 kpc of the centre of the Milky Way versus the same quantity for the LMC. Several points stand out clearly in the figures. First, HVS3 is the lone outlier as clearly preferring an LMC origin. Second, some of the candidates, such as HVS5, LAMOST-HVS1 to -HVS3, are much more consistent with a Galactic origin. Thirdly, the bulk of the population has a closer minimum approach to the Milky Way rather than the LMC, though here caution is warranted as the error bars remain large and there may be systematic uncertainties in the simulations.

\begin{figure*}
\centering
\includegraphics[width=0.33\textwidth]{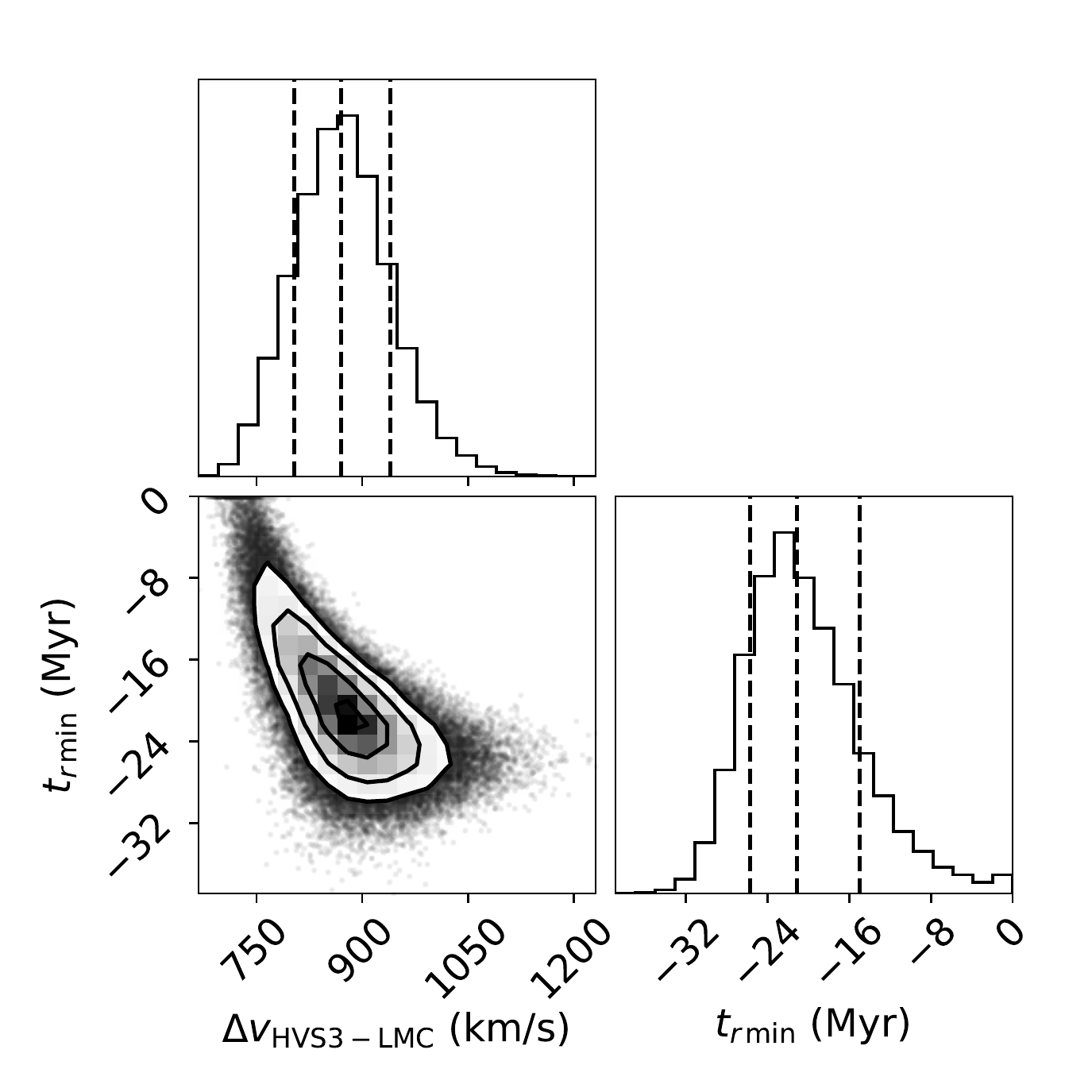}
\includegraphics[width=0.33\textwidth]{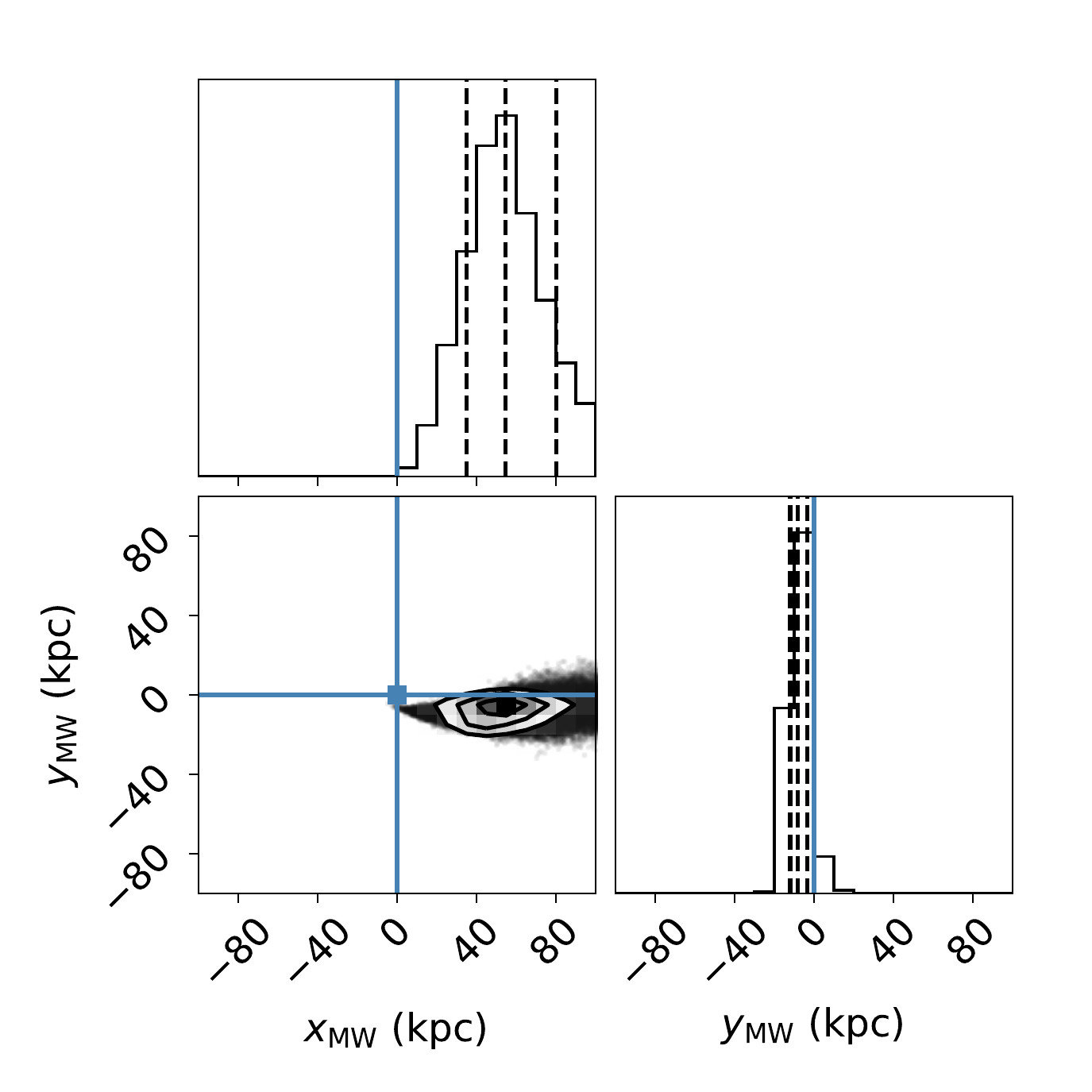}
\includegraphics[width=0.33\textwidth]{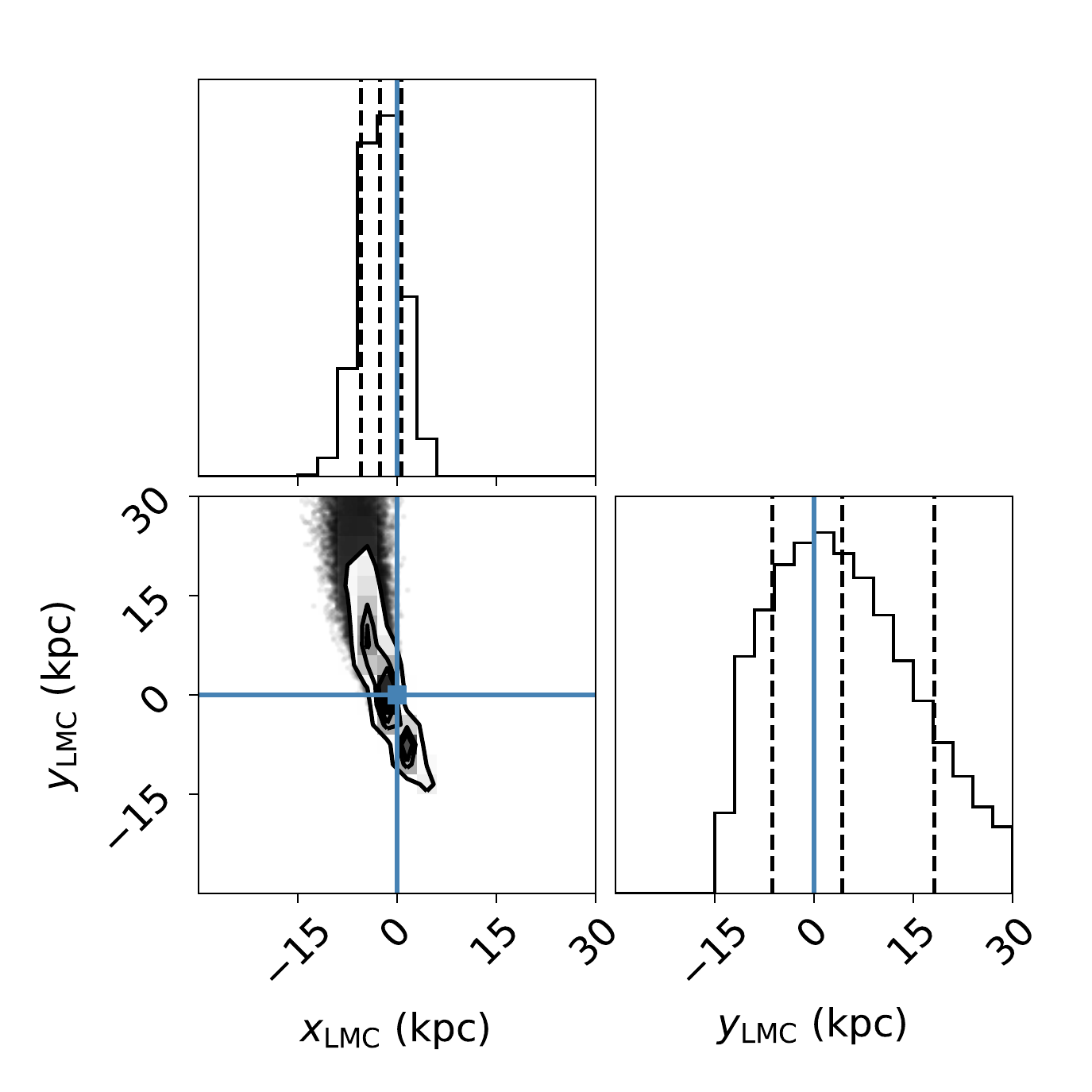}
\caption{\textbf{Left:} Distribution of relative velocities to the LMC and ejection times for HVS3. The dashed lines show the 15.9, 50, and 84.1 percentiles of each 1d distribution. The contours shown are 0.5, 1, 1.5. and 2$\sigma$. The ejection velocity is $870^{+69}_{-66}$ kms$^{-1}$. The ejection time is $21.1^{+6.1}_{-4.6}$ Myr (median with 1 sigma errors). 
\textbf{Middle:} Location where HVS3 passes through the plane of the Milky Way disk. The centre of the Milky Way excluded at the 3$\sigma$ level.  The centre of the Milky Way is a clear outlier. \textbf{Right:} Location where HVS3 passes through the plane of the LMC disk as defined in \protect\cite{vandermarel_lmc}. The centre of the LMC is just within the $1\sigma$ contour. We note that 8\% of the orbits for HVS3 do not pass through the plane of the LMC's disk.} 
\label{fig:HVS3}
\end{figure*}

\section{HVS3}
\label{sec:hvs3}

In \Figref{fig:HVS_MW}, HVS3 lies apart from the bulk of the population and
appears to pass very close to the centre of the LMC. In order to investigate it further, we sample its parameters (along with the LMC's parameters) 100,000 times and consider the resulting orbits. The panels of \Figref{fig:HVS3} show the results from these simulations. In the left panel, we show contours of the time of closest approach to the LMC and the relative velocity during this closest approach. The ejection velocity is $870^{+69}_{-66}$ kms$^{-1}$. The ejection time is $21.1^{+6.1}_{-4.6}$ Myr (median with 1 $\sigma$ errors). 

The middle panel of \Figref{fig:HVS_MW} shows the locations where HVS3 passes though the Galactic plane. Note that the Galactic Centre is excluded at the $3\sigma$ level. We find that the flight time to the point of closest approach with the Milky Way occurs $66^{+4.1}_{-4.5}$ Myr ago, significantly exceeding the main-sequence lifetime of 35 Myr for HVS3. Furthermore, this point of closest approach lies at a distance of $40^{+11}_{-10}$ kpc from the Milky Way centre.  The right panel of \Figref{fig:HVS3} shows where the orbits of HVS3 pass through the plane of the LMC disk \citep[as defined in][]{vandermarel_lmc}. This shows that HVS3 is consistent with being ejected from the very center of the LMC. This result was prefigured in the work of \citet{edelmann_he_2005} and \citet{gualandris_hvs3}, but the \textit{Gaia} DR2 proper motions have made the proposition unanswerable.

\subsection{Varying the Milky Way potential}

\begin{figure}
\centering
\includegraphics[width=0.45\textwidth]{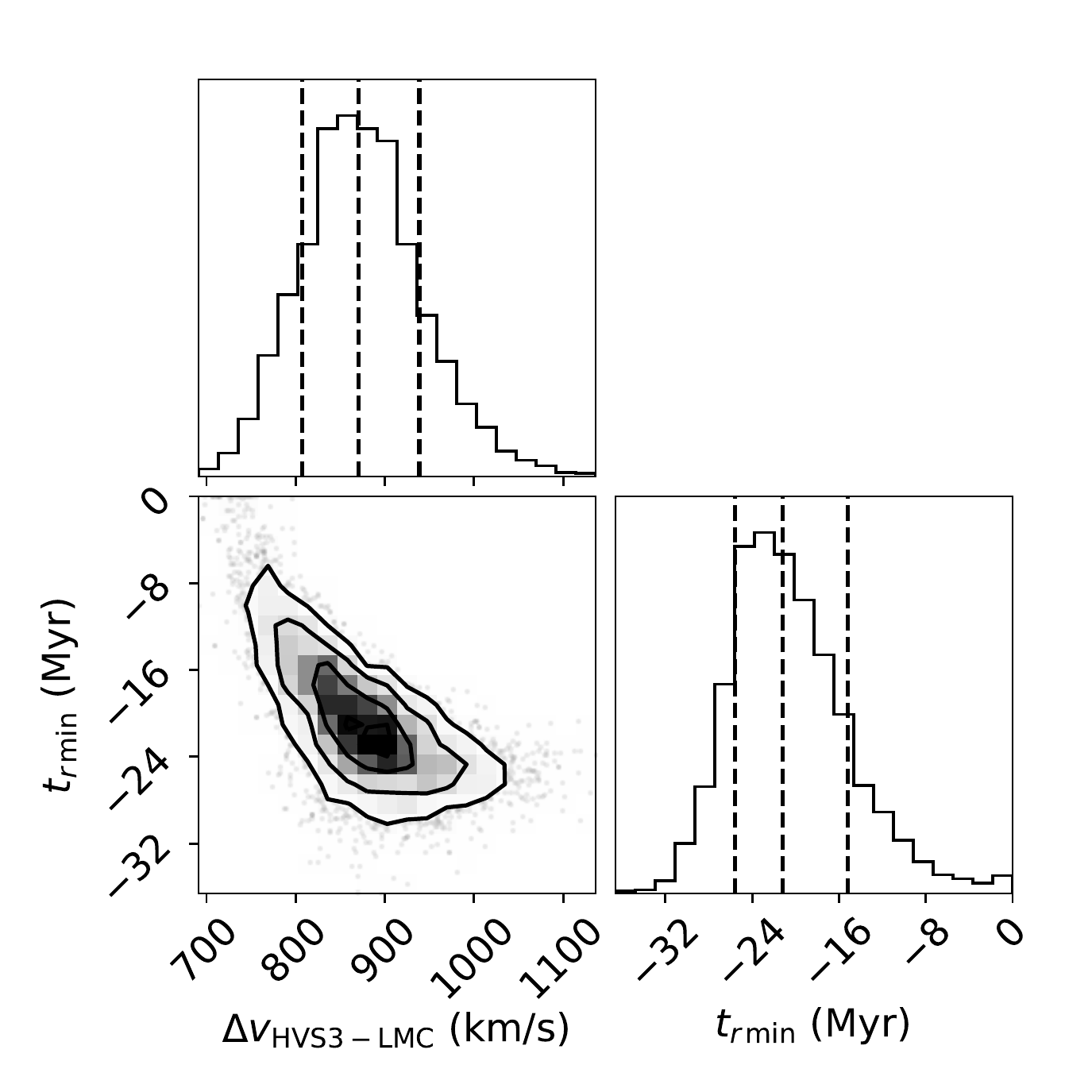}
\caption{Distribution of relative velocities to the LMC and ejection times for HVS3 evolved in the Milky Way potential from \protect\cite{pjm17}, marginalized over the potential uncertainties. The dashed lines show the 15.9, 50, and 84.1 percentiles of each 1d distribution. The ejection velocity is $871^{+68}_{-64}$ km/s. The ejection time is $21.2^{+6.0}_{-4.4}$ Myr (median with 1 sigma errors). The similarity of this result and that in \protect\figref{fig:HVS3} shows that the orbit of HVS3 is robust. } 
\label{fig:vrel_tej_HVS3_galpot}
\end{figure}

So far, we have assumed a single potential for the Milky Way, namely that of \texttt{MWPotential2014} from \cite{galpy}. In order to explore the effect of uncertainties in the Milky Way potential on the orbit of HVS3, we consider the fits to the Milky Way potential given in \cite{pjm17}. In addition to sampling the phase space coordinates of HVS3 and the LMC, we  sample the potential parameters fit in \cite{pjm17} the proper motion, distance, and radial velocity of HVS3 and the LMC, we also sample the potential parameters fit in \cite{pjm17}. These are the thin and thick disk surface density and scale radius, the bulge density, the NFW halo density and scale radius, and the solar position. These are sampled accounted for the correlation matrix given in \cite{pjm17}. The forces are implemented using \texttt{galpot} \citep{dehnen_binney_1998}. Finally, we also sample the Sun's motion relative to the local standard of rest (LSR) from the values given in \cite{pjm17}. This improved sampling thus represents a marginalization the uncertainties in the Milky Way potential, the Sun's location, and the Sun's velocity relative to the LSR. For HVS3, this produces a very similar result to what we found with the original potential. \Figref{fig:vrel_tej_HVS3_galpot} shows relative velocity and ejection time for HVS3 with respect to the LMC, which looks very similar to the left panel of \Figref{fig:HVS3} with an almost identical median and $1\sigma$ spread of the relative velocity and ejection time. 

Since the LMC is believed to have a mass on the order of $\sim 10^{11}M_\odot$, it is possible that the reflex motion of the Milky Way \citep[e.g.][]{gomez_et_al_2015} could have an effect on the origin of the HVSs. In order to test this, we treat the Milky Way as a particle sourcing a potential, \texttt{MWPotential2014} from \cite{galpy}. The rewinding procedure then integrates the orbit of the Milky Way, the LMC, and each HVS. This had a negligible effect on their trajectories and points of closest approach, likely due to the relatively short flight times for each HVS.

\subsection{Origin of HVS3}
The previous analysis strongly argues in favour of an origin in the LMC, as originally suggested by \citet{edelmann_he_2005}. The inferred relative velocity of $\sim 870 {\rm km\,s^{-1}}$ can be used to constrain the ejection mechanism. \citet{gualandris_three_body_2005} use  population synthesis of $(8+8)M_{\odot}$ binaries to show that the maximum escape velocity produced by binary evolution and supernova explosions is $\lesssim 100 {\rm km\,s^{-1}}$, even accounting for natal kicks.
It is also extremely unlikely that HVS3 is a binary system ejected as a result of a large asymmetric supernova kick, as the binary would have been unbound by the explosion.

The large observed velocity  therefore suggests an origin in a dynamical encounter in the LMC. Encounters with main-sequence stars, neutron stars or stellar mass black holes eject stars with maximum velocities of the order of the binary orbital velocity 
\citep[e.g][]{gualandris_three_body_2005}. Only a Hills type encounter with a massive black hole is able to produce a recoil velocity consistent with observations. In this case the ejection velocity is given by \citep{hills_hyper-velocity_1988}
\begin{equation}
V_{\rm ej} \sim 640\,{\rm km\,s^{-1}} \left(\frac{a_b}{0.2 {\rm AU}}\right)^{-1/2} \left(\frac{m_1+m_2}{16\,M_{\odot}}\right)^{1/3}\left(\frac{M_{\rm BH}}{10^3\,M_{\odot}}\right)^{1/6},
\end{equation}
where $a_b$ represents the binary semi-major axis, $m_1$ and $m_2$ the primary and secondary masses and $M_{\rm BH}$ the mass of the massive black hole. An ejection velocity of $871^{+68}_{-64}$ km/s thus requires a black hole mass of at least $4\times 10^3 - 10^4\,M_{\odot}$.
This is in good agreement with the early estimate of \citet{gualandris_hvs3} based on scattering experiments.

\section{Testing with mock simulations}\label{sec:testing}

In this section we will use two N-body simulations with hypervelocity stars ejected from the Milky Way and LMC centres to test how well we can recover their properties, as well as estimate the probability of misclassifying a star.

The simulations used in this work are described in \cite{boubert_hvs_runaway}. These consist of N-body simulations evolved with a modified version of \textsc{gadget-3},  which is an improved version of \textsc{gadget-2} \citep{gadget2}. The code has been modified so that we can track the centre of either the Milky Way or the LMC on the fly. It has also been modified to eject massless tracer particles (representing HVSs) from the centre of either galaxy.

The simulations \citep[see][for more details]{boubert_hvs_runaway} evolve an N-body Milky Way (bulge, disk, and halo) and an N-body LMC (disk and halo) which are initialized so that the LMC ends up within $2\sigma$ of its observed location and velocity at the end of the simulation. We discuss two separate simulations below, one where HVSs are ejected from the centre of the LMC and one where they are ejected from the Milky Way centre.

\subsection{Tests with stars ejected from the LMC}

In this section, we take a mock sample of hypervelocity stars from  \cite{boubert_hvs_runaway} which were ejected by a massive black hole in the centre of the LMC. These were ejected with a velocity distribution governed by the Hills mechanism, assuming a black hole mass of $1.7\times 10^5 M_\odot$ in the centre of the LMC \citep[see][for more details]{boubert_dipole_2016}. The simulation was evolved for 350 Myr which is sufficient to model the young stars ejected from the LMC. 

For each HVS in our sample, we locate the closest particle to the 3d location of the HVS (ignoring distance uncertainties). This gives us a sample of 26 mock HVSs which were ejected from the LMC and are close to the observed HVSs. We note that almost all of the observed HVSs had analogues within 5 kpc of their present day position. The distance, proper motion, and radial velocity of each mock star is then perturbed by the observational errors of its associated HVS. For each of these mock HVSs, we then repeat the same rewinding procedure described in \Secref{sec:population} to determine how well we can identify the origin of these stars. We note that the rewinding procedure was identical to what we did on the observed sample of HVSs, except we changed the mass of the Milky Way's NFW to be $10^{12} M_\odot$, equal to the simulated Milky Way's halo mass.

\begin{figure}
\centering
\includegraphics[width=0.45\textwidth]{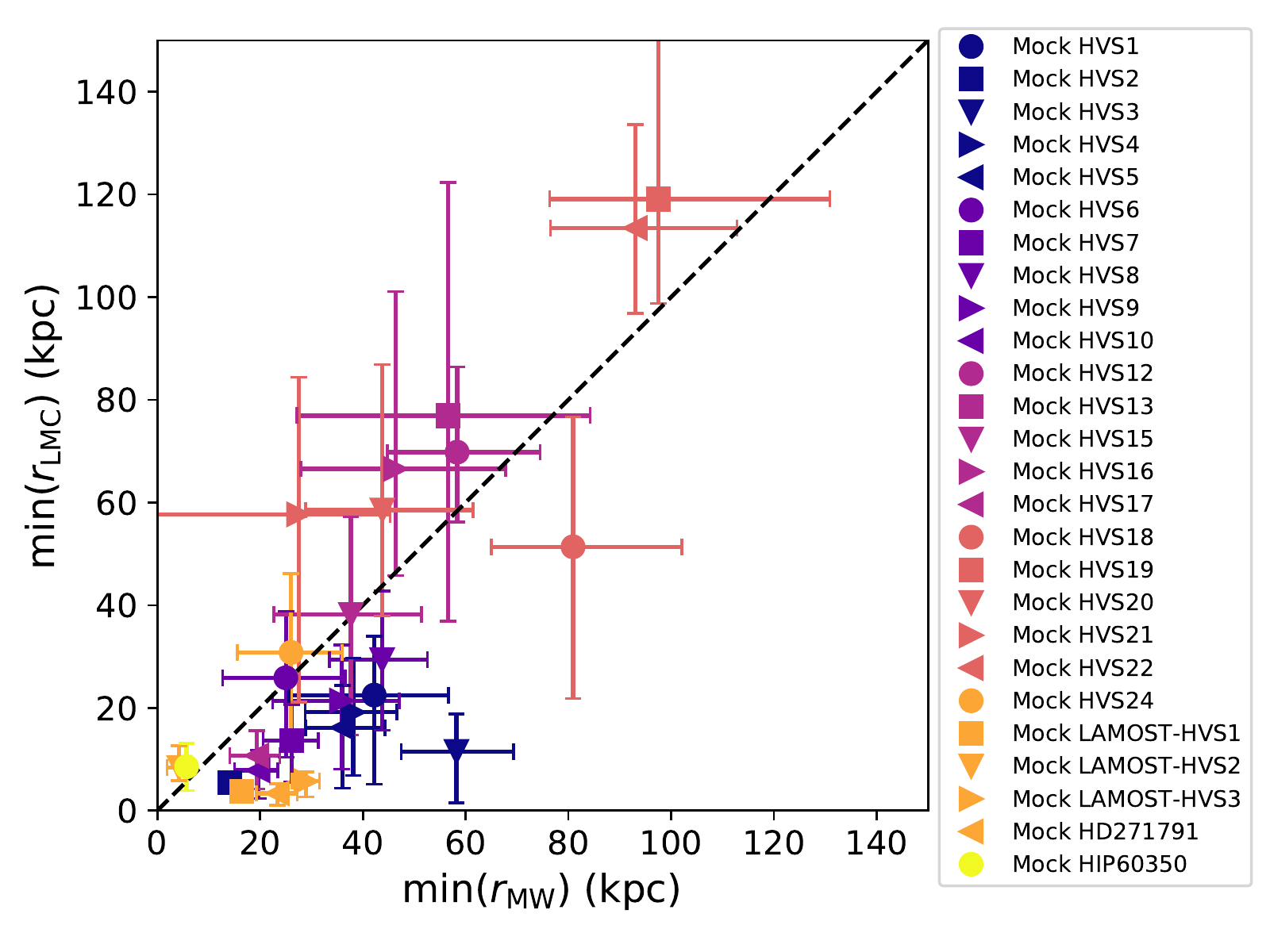}
\caption{Minimum approach distance to the Milky Way and the LMC for a mock sample of hypervelocity stars. } 
\label{fig:rmin_HVS_mock}
\end{figure}

\Figref{fig:rmin_HVS_mock} shows the distance of closest approach to the Milky Way and the LMC for this mock sample. While many of the mock HVSs are consistent with coming from the LMC (i.e. the clustering of stars near the x-axis of the figure), there are many stars which appear inconsistent with an LMC origin. Many of the stars are distant from both the Milky Way and the LMC (similar to the observed HVSs, left panel of \figref{fig:HVS_MW}). This suggests that there is a significant possibility that other stars in this sample of 26 could have originated in the LMC. Improved data is needed to determine their origin. However, this figure is clearly different from the left panel of \Figref{fig:HVS_MW} which shows that only one of the observed HVSs comes from the LMC. Thus, it appears that the majority of the observed sample did not come from the LMC.

\subsection{Tests with stars ejected from the Milky Way}

Next, we take mock stars from simulations which include a population of HVSs ejected from the Milky Way centre. We use this simulation to assess how often we misclassify a HVS ejected from the Milky Way as coming from the LMC. These simulations are almost identical to the ones described in the previous section except the HVSs are ejected from the Milky Way centre instead of from the LMC. The stars are ejected with a velocity distribution similar to that in \cite{boubert_hvs_runaway} except with a black hole mass of $4\times 10^6M_\odot$ \citep{ghez_etal_2005}. The simulation was evolved for 1 Gyr.

We create a mock sample by sampling from the distance to HVS3 and then selecting the closest star in the simulation (in the direction towards HVS3) which has not already been selected. This gives us a sample of 1000 mock stars drawn from a similar location to HVS3. We note that we do not make any selection on the velocities of the particles in the simulation. 

Each particle is then assigned the observational errors of HVS3 (in distance, proper motions, and radial velocity) and a mock observation is performed, perturbing the particles properties from their true values. These values represent the mock star's properties and we sample from its errors as in \Secref{sec:population}. For this procedure, we consider two sets of errors for the proper motion: HST errors from \cite{brown_proper_2015} and the proper motion errors from \textit{Gaia} DR2. This comparison is instructive since it allows us to see the resolving power of \textit{Gaia}. 

\begin{figure}
\centering
\includegraphics[width=0.45\textwidth]{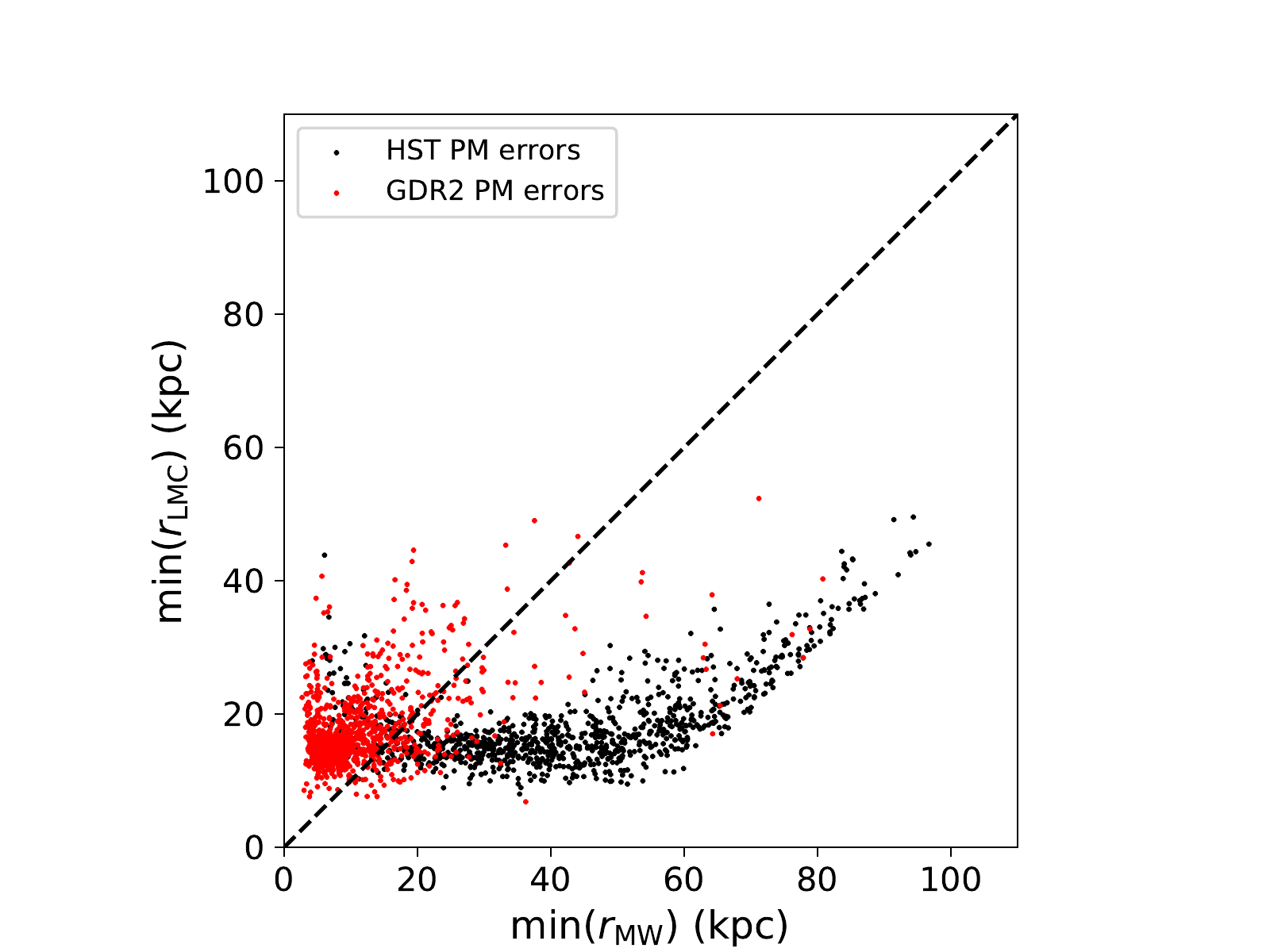}
\caption{Comparison of minimum approach distance to the MW and LMC for 1000 mock stars ejected from the Milky Way chosen at the location of HVS3. We see that with HST proper motions, many of the mocks appear to have passed closer to the MW than the LMC. However, this is a bias due to the large errors with HST. With GDR2 errors, almost all of the stars are correctly inferred to pass much closer to the Milky Way than the LMC. Thus, the proper motion quality of GDR2 is critical to correctly determine whether HVS3 originated in the Milky Way or the LMC.} 
\label{fig:HVS3_MW_BH_mock}
\end{figure}

\begin{figure}
\centering
\includegraphics[width=0.45\textwidth]{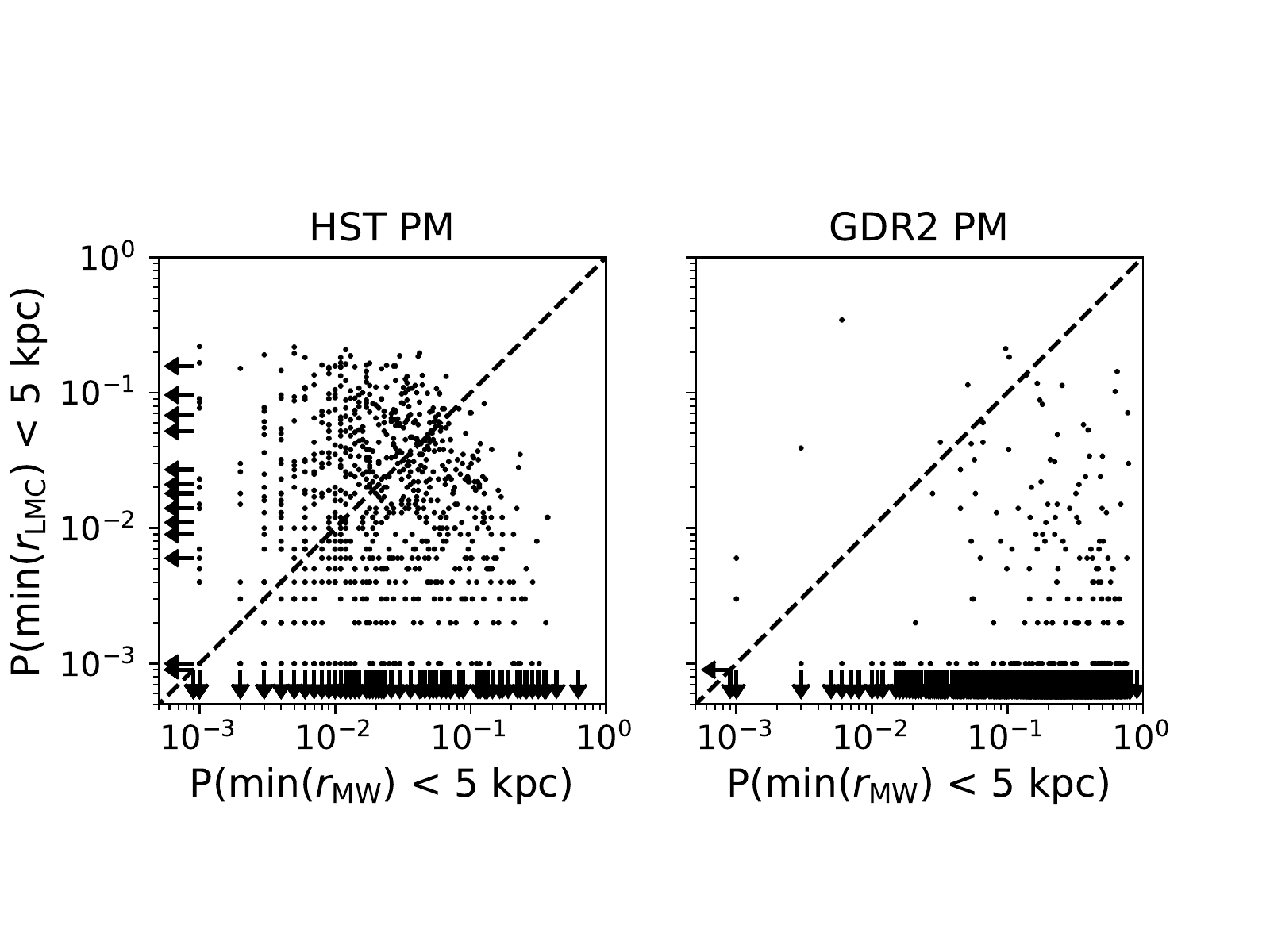}
\caption{Comparison of the probability of approaching within 5 kpc for 1000 mock stars ejected from the Milky Way selected at the location of HVS3. This figure shows that the false positive rate of misclassifying a MW hypervelocity star as coming from the LMC is approximately 1 a 1000. } 
\label{fig:pMW_pLMC_HVS3_MW_BH_mock}
\end{figure}

\Figref{fig:HVS3_MW_BH_mock} shows the closest approach distance to the Milky and the LMC for the sample of mock stars. With HST quality proper motions \citep[based on the uncertainties for HVS3 in][]{brown_proper_2015}, there is a significant risk of misclassifying a Milky Way HVS as coming from the LMC. This bias is due to the fact that HVS3 is closer to the LMC than the Milky Way and thus there is a larger spread in the orbits by the time they reach the Milky Way. However, if we use the proper motions from \textit{Gaia} DR2, we find a significant improvement with few Milky Way HVS appearing to originate in the LMC. \Figref{fig:pMW_pLMC_HVS3_MW_BH_mock} shows the probability of these mock HVSs passing within 5 kpc of either the Milky Way of the LMC. This shows that \textit{Gaia} DR2 quality observations were needed to reduce the risk of misclassifying a Milky Way HVS as coming from the LMC. Based on this figure we estimate that the chance that HVS3 is misclassified is approximately 1 in 1000. 

\section{Conclusion}

We present a first look at the orbits of 26 hypervelocity stars around the Milky Way using proper motions from the recent \textit{Gaia} DR2. These exquisite proper motions allow us to determine the origin of a number of these HVSs. While many of them appear to come from the Milky Way, HVS3 is a clear outlier and comes from the LMC, thus confirming the original suggestions of \cite{edelmann_he_2005} and \cite{gualandris_hvs3}. We find that HVS3 had a closest approach with respect to the LMC $21.1^{+6.1}_{-4.6}$ Myr ago, with a relative velocity of $870^{+69}_{-66}$ km s$^{-1}$, and is consistent with being ejected from the very centre of the LMC. This high ejection velocity rules out all scenarios except for the Hills mechanism, requiring a black hole mass of at least $4\times 10^3 -10^4 M_\odot$. This provides strong direct evidence that the LMC harbours a massive black hole. 

We run a number of tests to ensure that this result is robust. We have marginalized over the Milky Way potential from \cite{pjm17}, which has a negligible effect on our results. We also run our analysis on two N-body simulations which contain HVSs from the LMC and the MW. The first test with mock HVSs from the LMC showed that if all of the 26 HVSs in our sample originated from the LMC, we should have found more HVSs with a close LMC passage. This tentatively suggests that at least some of the 26 HVS did not originate in the LMC. Interestingly, this test also revealed that a number of the poorly constrained HVSs are consistent with an LMC origin. The second test with a mock of Milky Way HVSs was only performed for HVS3. This showed that \textit{Gaia} DR2 quality proper motions were needed to confirm whether HVS3 indeed originated in the LMC, and thus to provide strong evidence of a massive black hole in the LMC's centre.

\section*{Acknowledgements}

D. Boubert thanks the UK Science and Technology Facilities Council for supporting his PhD. This work has made use of data from the European Space Agency (ESA) mission {\it Gaia}(\url{https://www.cosmos.esa.int/gaia}), processed by the {\it Gaia} Data Processing and Analysis Consortium (DPAC, \url{https://www.cosmos.esa.int/web/gaia/dpac/consortium}). Funding for the DPAC has been provided by national institutions, in particular the institutions participating in the {\it Gaia} Multilateral Agreement.

\bibliographystyle{mn2e_long}
\bibliography{citations_hvs}

\appendix

\section{Properties of HVS} \label{sec:hvs_props}

We list the properties of the 26 HVSs studied in this work in Table \ref{tab:hvs}.

\begin{table*}[]
\centering
\caption{The 26 hypervelocity stars considered in this work. Almost all of the proper motion measurements are from \textit{Gaia} DR2 except for HVS1, HVS4, HVS13, and HIP60350. {\scriptsize \textbf{References:} (1) \citet{brown_proper_2015} (2) \citet{2018arXiv180409365G}, (3) \citet{geier_fastest_2015}, (4) \citet{brown_mmt_2014}, (5) \citet{zheng_first_2014}, (6) \citet{huang_discovery_2017}, (7) \citet{heber_hypervelocity_2008}, (8) \citet{irrgang_nature_2010}.}}
\label{tab:hvs}
\begin{tabular}{lllrrrrrc}
\hline ID          & R.A.        & Dec.         & $\mu_{\alpha}\cos{\delta}\;(\mathrm{mas}\;\mathrm{yr}^{-1})$                 & $\mu_{\mathrm{\delta}}\;(\mathrm{mas}\;\mathrm{yr}^{-1})$  & $\mathrm{Corr}(\mu_{\alpha}\cos{\delta},\mu_{\mathrm{\delta}})$ & $v_{\mathrm{hel}}\;(\mathrm{km}\;\mathrm{s}^{-1})$   & $D_{\mathrm{hel}}\;(\mathrm{kpc})$  & Ref.  \\ \hline
HVS1        & 09:07:44.99 & +02:45:06.9  & $0.1    \pm 0.3$     & $-0.1   \pm 0.2  $    & ...               & $831.1  \pm 5.7 $    & $107   \pm 15  $    & 1     \\
HVS2        & 09:33:20.86 & +44:17:05.4  & $-5.363 \pm 0.391$   & $1.285  \pm 0.382$    & -0.2092           & $917    \pm 7   $    & $8.5   \pm 1   $    & 1,2,3 \\
HVS3        & 04:38:12.8  & -54:33:12    & $0.851  \pm 0.110$   & $1.936  \pm 0.162$    & 0.1899            & $723    \pm 3   $    & $61    \pm 10  $    & 1,2   \\
HVS4        & 09:13:01.01 & +30:51:19.8  & $-0.2   \pm 0.4 $    & $-0.4   \pm 0.4  $    & ...               & $600.9  \pm 6.2 $    & $64    \pm 9.8 $    & 1     \\
HVS5        & 09:17:59.48 & +67:22:38.3  & $-0.023 \pm 0.176$   & $-1.179 \pm 0.268$    & 0.2162            & $545.5  \pm 4.3 $    & $45    \pm 5.2 $    & 1,2   \\
HVS6        & 11:05:57.45 & +09:34:39.47 & $-0.378 \pm 0.664$   & $-0.503 \pm 0.507$    & -0.0242           & $609.4  \pm 6.8 $    & $55    \pm 6.9 $    & 1,2   \\
HVS7        & 11:33:12.12 & +01:08:24.9  & $-0.677 \pm 0.373$   & $0.457  \pm 0.253$    & -0.2767           & $526.9  \pm 3   $    & $52    \pm 6.4 $    & 1,2   \\
HVS8        & 09:42:14.04 & +20:03:22.1  & $-0.972 \pm 0.365$   & $0.117  \pm 0.369$    & -0.3844           & $499.3  \pm 2.9$     & $53    \pm 9.8 $    & 1,2   \\
HVS9        & 10:21:37.08 & -00:52:34.8  & $0.412  \pm 0.743$   & $-0.213 \pm 0.747$    & -0.2503           & $616.8  \pm 5.1 $    & $74    \pm 12  $    & 1,2   \\
HVS10       & 12:03:37.85 & +18:02:50.4  & $-2.597 \pm 1.288$   & $-0.788 \pm 0.494$    & 0.0151            & $467.9  \pm 5.6 $    & $52    \pm 5.8 $    & 1,2   \\
HVS12       & 10:50:09.59 & +03:15:50.67 & $0.42   \pm 1.377$   & $0.285  \pm 0.993$    & 0.2614            & $552.2  \pm 6.6 $    & $66    \pm 8.5 $    & 1,2   \\
HVS13       & 10:52:48.3  & -00:01:33.94 & $-0.9   \pm 0.4  $   & $0.5    \pm 0.4  $    & ...               & $569.3  \pm 6.1 $    & $105   \pm 19  $    & 1     \\
HVS15       & 11:33:41.09 & -01:21:14.25 & $-0.498 \pm 1.291$   & $-0.487 \pm 0.567$    & -0.0546           & $461    \pm 6.3 $    & $67    \pm 10  $    & 2,4   \\
HVS16       & 12:25:23.4  & +05:22:33.84 & $-1.895 \pm 1.518$   & $-1.145 \pm 0.856$    & -0.429            & $429.8  \pm 7   $    & $71    \pm 12  $    & 2,4   \\
HVS17       & 16:41:56.39 & +47:23:46.12 & $-1.034 \pm 0.198$   & $-1.065 \pm 0.323$    & 0.0801            & $250.2  \pm 2.9 $    & $49    \pm 4   $    & 2,4   \\
HVS18       & 23:29:04.94 & +33:00:11.47 & $-0.123 \pm 0.656$   & $-0.119 \pm 0.495$    & 0.213             & $237.3  \pm 6.4 $    & $80    \pm 11  $    & 2,4   \\
HVS19       & 11:35:17.75 & +08:02:01.49 & $-0.295 \pm 1.790$   & $0.413  \pm 1.224$    & -0.3625           & $592.8  \pm 11.8$    & $98    \pm 15  $    & 2,4   \\
HVS20       & 11:36:37.13 & +03:31:06.84 & $0.453  \pm 1.451$   & $0.569  \pm 1.014$    & -0.2955           & $512.1  \pm 8.5 $    & $76    \pm 11  $    & 2,4   \\
HVS21       & 10:34:18.25 & +48:11:34.57 & $0.099  \pm 0.693$   & $-0.263 \pm 0.881$    & -0.3749           & $356.8  \pm 7.5 $    & $113   \pm 21  $    & 2,4   \\
HVS22       & 11:41:46.44 & +04:42:17.29 & $0.184  \pm 2.024$   & $1.987  \pm 1.443$    & -0.2758           & $597.8  \pm 13.4$    & $85    \pm 13  $    & 2,4   \\
HVS24       & 11:11:36.44 & +00:58:56.44 & $0.301  \pm 0.777$   & $-0.408 \pm 0.578$    & -0.1616           & $492.5  \pm 5.3 $    & $56    \pm 7   $    & 2,4   \\
LAMOST-HVS1 & 09:12:06.52 & +09:16:21.8  & $-3.537 \pm 0.110$   & $-0.62  \pm 0.093$    & -0.2516           & $611.65 \pm 4.63$    & $13.91 \pm 2   $    & 2,5   \\
LAMOST-HVS2 & 16:20:20.77 & +37:47:39.9  & $-2.563 \pm 0.056$   & $-0.924 \pm 0.072$    & 0.2943            & $341.1  \pm 7.79$    & $22.24 \pm 4.6 $    & 2,6   \\
LAMOST-HVS3 & 03:21:17.08 & +19:07:36.2  & $1.509  \pm 0.199$   & $-0.281\pm 0.155$     & 0.0972            & $361.38 \pm 12.52$   & $22.32 \pm 2.5 $    & 2,6   \\
HD271791    & 06:02:27.88 & -66:47:28.68 & $-0.619 \pm 0.067$   & $4.731  \pm 0.071 $   & 0.0996            & $441    \pm 1    $   & $21    \pm 4   $    & 2,7   \\
HIP60350    & 12:24:30.23 & +34:57:58.72 & $-13.51 \pm 1.31 $   & $16.34  \pm 1.37$     &  ...              & $262    \pm 5    $   & $3.1   \pm 0.6 $    & 8    \\ \hline
\end{tabular}
\end{table*} \label{tab:hvs}

\end{document}